\documentclass[preprint,showpacs,preprintnumbers,amsmath,amssymb]{revtex4}

\usepackage{graphicx}
\usepackage{dcolumn}
\usepackage{bm}

\begin{document}

\title { Magnetization reversal driven by spin-injection :

a mesoscopic spin-transfer effect}
\author{J.-E. Wegrowe} \email{jean-eric.wegrowe@polytechnique.edu}
\author{S. M. Santos,  M.-C.
Ciornei, H.-J. Drouhin}
\affiliation{Laboratoire des Solides Irradi\'es,
Ecole Polytechnique, 91128
Palaiseau Cedex, France.} 
\author{J. M. Rub\'i}
\affiliation{Departement de Fisica Fonamental,
Universitat de Barcelona, Diagonal 647, Barcelona 08028,
Spain.} 

\date{\today}

\begin{abstract}
A mesoscopic description of spin-transfer effect is proposed, based
on the spin-injection mechanism occurring at the junction with
a ferromagnet. The effect of spin-injection is to modify locally, in
the ferromagnetic configuration space, the density of magnetic
moments.  The corresponding gradient leads to a current-dependent
diffusion process of the magnetization.  In order to describe this
effect, the dynamics of the magnetization of a ferromagnetic single
domain is reconsidered in the framework of the thermokinetic theory of
mesoscopic systems.
Assuming an Onsager cross-coefficient that couples the currents, it is
shown that spin-dependent electric transport leads to a correction of
the Landau-Lifshitz-Gilbert equation of the ferromagnetic order
parameter with supplementary diffusion terms.  The consequence of
spin-injection in terms of activation process of the ferromagnet is
deduced, and the expressions of the effective energy barrier and of
the critical current are derived.  Magnetic fluctuations are
calculated: the correction to the fluctuations is similar to that
predicted for the activation.  These predictions are consistent with
the measurements of spin-transfer obtained in the activation regime
and for ferromagnetic resonance under spin-injection.

\end{abstract}

\pacs{75.40.Gb,72.25.Hg,75.47.De \hfill}

\maketitle

In the context of spintronics, spin transfer is a generic term that
describes the magnetization reversal or magnetization excitations of a
ferromagnetic layer provoked by the injection of a spin-polarized
electric current.  This effect has now been observed while measuring
the magnetization of many different systems \cite{liste0}.  It has
also been implemented into the last generation of magnetic random
access memories \cite{MRAM}.

Spintronics immerged first with the discovery of spin-injection 
and giant magnetoresistance (GMR) \cite{GMR}.
Effects of spin-injection at a junction of a ferromagnet are
successfully described within the two spin-channel model
\cite{Johnson,Wyder,ValetFert,Levy,Weg}.  In this description,
GMR, or spin-accumulation, is due to the spins
of the conduction electrons that are driven out-of equilibrium by an
interface: the difference of the electro-chemical potentials $\Delta
\mu$ between the two spin-channels leads to a redistribution of the
spin populations through spin-flip mechanisms.  Depending on the
magnetic state of the ferromagnet at the junction, this redistribution
of spin-dependent electronic populations modifies the resistance.  
For this reason, the GMR can be used as a probe (e.g. read heads for 
hard disk drive) in order to measure
precisely the position of a nanoscopic ferromagnetic moment in the
corresponding configuration space.  Inversely, from the point of view of
the ferromagnetic properties, it is natural to expect that the
spin-flip mechanism at the interface modifies locally the density of the
magnetic moments in this configuration space.  This is a consequence
of both the spin-orbit coupling (or s-d relaxation), and the
redistribution of spins in the magnetic configuration space.  The
spin-injection would then be responsible for a gradient of the density
in the magnetic configuration space, that contributes to the diffusion
processes of the ferromagnetic moment.

The aim of this work is to investigate the consequences of this
redistribution mechanism for the ferromagnetic moments. 
This task is performed in the framework of the nonequilibrium theory
of mesoscopic systems \cite{Rubi1,RubiFDT}.  In this context, the
coupling between the spin of conduction electrons and the
ferromagnetic order parameter is due to the introduction of a relevant
Onsager cross-coefficient that links the spin-polarized electric
currents (i.e. transport of mass, spins, and electric charges in the
normal space) to the ferromagnetic current (i.e. a massless transport
of ferromagnetic moments in the corresponding space).  Physically,
this spin-transfer cross-coefficient accounts for the fact that the
spins of the conduction electrons also contribute to the transport of
the ferromagnetic order parameter in the space of magnetic moments, in
analogy with the thermoelectric Peltier-Seebeck cross-coefficient that
accounts for the fact that charge carriers contribute also to the
transport of heat.

The motivation for a classical mesoscopic analysis was the need to account for
the following highly specific properties observed in spin-transfer
experiments that can hardly be accounted for in direct microscopic approaches
\cite{Slon,Berger,Miltat,Revue}.  First, in single domain
ferromagnets, the reversible part of the hysteresis loop is not
significantly modified under current injection, while the irreversible
jump is drastically modified \cite{Revue}.  Second, the amplitude of
spin-transfer is proportional to the giant magnetoresistance and to
the amplitude of the current \cite{MSU2,Jiang,Manchon}.  Third, the
N\'eel-Brown activation law is still valid under current injection,
with a correction of the barrier height that is quasi-symmetric under
both the permutation of the magnetic configuration (parallel to
anti-parallel and inversely) and the change of the direction of the
current \cite{MSU2,Moi,Fabian,Buhrman}.  "Quasi"-symmetric means here
that a quantitative shift (a factor 2 to 4 in general) is
systematically observed in the amplitude of spin transfer for both
transitions in spin-valve structures.  Fourth, this last
quasi-symmetry is also observed in the context of ferromagnetic
resonance under current injection close to the equilibrium states
(i.e. with an weak spin-injection) \cite{DienyRes}.  In order to
compare the results of the model with these observations, after
calculating the diffusive correction of the Landau-Lifshitz-Gilbert
equation (LLG) due to spin-injection in the first section, the
correction to activation law is derived in a second section, and the
correction to the fluctuations is derived in the third section.

 \subsection{Spin-injection correction to the ferromagnetic
 Landau-Lifshitz-Gilbert Equation}

In the framework of the two conducting channel approximation, the
system is described by two electronic populations.  A first conducting
channel is carrying the conduction electrons of spin up $\uparrow$
with the electric conductivity $\sigma_{\uparrow}$ and the other
channel is carrying the electron conduction of spin down $\downarrow $
with the electric conductivity $\sigma_{\downarrow}$.  The
quantification axis is defined by the direction of the magnetization
of the ferromagnet.  The Ohm's law is valid for each channel: in a 1D
ferromagnetic wire described with a single space coordinate $z$, the
electric current $J^{e}_{\uparrow}$ (resp.  $J^{e}_{\downarrow}$ ) is
related to the {\it local} electric field $E_{\updownarrow}=
-\frac{1}{e} \frac{\partial \mu^{e}_{\updownarrow}}{ \partial z}$
\cite{Levy} through the conductivity: $J^{e}_{\updownarrow} =
\sigma_{\updownarrow} E_{\updownarrow}$, where $\mu^{e}_{\uparrow}$
(resp.  $\mu^{e}_{\downarrow}$) is the electrochemical potential of
the channel of spins $\uparrow$ (resp.  $\downarrow$).

In the case of an interface between a normal metal and a ferromagnet
(or between two ferromagnets), the spin-dependent relaxation between
both electronic populations leads to a redistribution of spins within
the two channels, that is driven by the difference between the two
electro-chemical potentials $\Delta \mu^{e} (z) = \mu^{e}_{\uparrow} -
\mu^{e}_{\downarrow}$ (the "spin-neutral" electrochemical potential is
defined, in turns, by $\mu^{e}_{0} = \mu^{e}_{\uparrow} +
\mu^{e}_{\downarrow}$).  The parameter $\Delta \mu^{e}$ accounts for
the spin-injection, or spin accumulation, mechanism (see Fig.  1). 
The equilibrium state of the spin system is recovered in the bulk
where, by definition, $\Delta \mu^{e} (\infty) = 0$.  This corresponds
to a distance of some few times the spin-diffusion length (some tens
of nanometers in usual Co or Ni layers).  The description can be
generalized to multichannel model that includes the spins of the
conduction electron of the $s$ band and of the spins of the conduction
electrons of the $d$ band, with the corresponding interband relaxation
\cite{MTEPW,Revue}.  For convenience, a two-channel approximation is
used in the following, in which we define $J_{0}^e = J^{e}_{\uparrow}
+ J^{e}_{\downarrow}$ as the "spin-neutral" electrical current, and
$\delta J^e = J^{e}_{\uparrow} - J^{e}_{\downarrow}$ is the
spin-polarized electrical current.  \\

The system under consideration is composed not only by the microscopic
spins up and down carried by the conduction electrons of different
nature (band $s$ or $d$), but also by a ferromagnet of length $v$ (
$v$ is also the volume in the normal space for a section unity)
described by a ferromagnetic order parameter $\vec M$.  The last
variable is defined in the space of the ferromagnetic moments (
Fig.  2) of constant modulus : $\vec M = M_{s} \vec u_{r}$, called
$\gamma$ - space in the following \cite{Prigogine}.  This space can be defined on the
unit sphere (see Fig.  2) with the two angles $\theta$ and $\varphi$,
the radial unit vector $\vec u_{r}$, the azimuth unit vector $\vec
u_{\theta}$ and the zenith unit vector $\vec u_{\varphi}$ .  The
magnetization is then described statistically in the configuration
space by the density $\rho^{F}(\theta, \varphi)$ of ferromagnetic
moments oriented at a given direction $\gamma = \{\theta,\varphi \} $,
and also by the ferromagnetic potential energy $V^{F}(\theta, \varphi)$ that
contains at least the contributions due to the external magnetic field 
$\vec{H}$
and the anisotropy energy. Typically, for a uniaxial anisotropy with 
anisotropy constant $K$, the ferromagnetic potential writes:  $V^{F}(\theta, \varphi) = 
K sin^{2}(\theta) - M_{s} H cos(\theta - \phi)$ where $\phi$ gives the 
direction of the applied field. This potential energy has the form of a double 
well potential (Fig.  2).

In the absence of spin-injection, the magnetization is a conserved
variable, so that the conservation law writes $\frac{\partial
\rho_{0}^{F}}{\partial t} = - div(\vec{j}_{0}^{F}) $, where
$\vec{j}_{0}^F = j_{0}^{F \theta} \vec{u}_{\theta} + j_{0}^{F \varphi}
\vec{u_{\varphi}}$ is the ferromagnetic current density and the
operator $div$ is the divergence defined on the surface of a unit
sphere.  This is no longer the case under spin-injection at an
interface: due to the redistribution of spins in the different
channels (especially from $s$ band to $d$ band) spins are transferred
from one sub-system to the other, and the ferromagnetic sub-system becomes
an {\it open system}.  In order to work in a larger system that is closed, i.e. 
that does not exchange magnetic moments with the environment, the total
ferromagnetic density $\rho^{F}_{tot}$ and total ferromagnetic current
$\vec{j}_{tot}^F $ are defined in what follows.

In the total system, that includes both the ferromagnetic layer and
the spin-polarized current, the entropy production d$\mathcal{S}$/dt
(per unit of solid angle and per unit of length) is given by

\begin{equation}
\frac{d\mathcal{S}}{dt}= - \frac{1}{T} \left ( \vec{j}_{tot}^F . \vec {\nabla} \mu
^F - \delta J^e \frac {\partial{\Delta {\mu}^e}} {e \partial {z}} -
J_{0}^e \frac { \partial{\mu_{0}^e}} {e \partial {z}} \right )
\label{source}
\end{equation}
where $T$ is the temperature assumed uniform,
$\vec{\nabla}$ is the gradient defined on the surface of the
unit sphere, $\mu ^F$ is the total ferromagnetic chemical potential, and $e $
the charge of the electron. The 
last term in the right hand side is the Joule heating, the second term is 
the dissipation related to the giant magnetoresistance, and the 
first term is the ferromagnetic dissipation that {\it defines} the 
total ferromagnetic current $\vec{j}_{tot}^F$ in the internal space of magnetic 
moments \cite{DeGroot}.

 From the expression of the entropy production Eq.  (\ref{source}) and
 the second law of thermodynamics d$\mathcal{S}/dt \ge 0$, the flux
 involved in the system are related to the generalized forces through the
 matrix of the Onsager transport coefficients \cite{Stueckelberg}

\begin{equation}
\left( \begin{array}{c}
j^{\varphi F} \\
j^{\theta F} \\
\delta J^e \\
J^{e}_{0}
\end{array} \right)
=  - \left( \begin{array}{cccc}
                        L_{\varphi \varphi} & L_{ 
                        \varphi 
						\theta} & 0 & 0 \\
                        L_{\theta \varphi} & L_{\theta 
                        \theta} & l & 0 \\
                       0 & \tilde l & \sigma_0 & \beta \sigma_0 \\
                        0 & 0 & \beta \sigma_0 & 
                        \sigma_0
                        \end{array} \right)
                        \left( \begin{array}{c}
 \frac{1}{sin( \theta)} \frac{\partial \mu ^F}{\partial \varphi}\\
 \frac{\partial \mu ^F}{\partial \theta}\\
\frac{1}{e} \frac {\partial{\Delta 
{\mu}^{e}}}{\partial {z}}\\
\frac{1}{e} \frac{\partial \mu^{e}_{0}}{\partial z}
 \end{array} \right)
\label{Onsagermatrix}
\end{equation}

All coefficients are known, except the new cross-coefficients $l$ and
$\tilde l$, introduced in this model, and related to the experimental
parameters at the end of the next section.  The electric conductivity
$\sigma_{0}$ is given, in the two channel approximation, by
$\sigma_{0} = \frac{\sigma_{\uparrow} + \sigma_{\downarrow}}{2} $, and
the conductivity asymmetry $\beta$ is given by $\beta =
\frac{\sigma_{\uparrow} - \sigma_{\downarrow}}{\sigma_{0}}$.  The four
ferromagnetic transport coefficients are $ \{ L_{\theta\theta}, L_{\theta \varphi},
L_{\varphi \theta}, L_{\varphi \varphi} \}$. 
However, the Onsager-Casimir reciprocity relations give $L_{\theta
\varphi}= - L_{\varphi \theta}$, and the symmetry imposes that
$L_{\theta \theta}=L_{\varphi \varphi}$ \cite{Revue}.  Furthermore,
the two transport coefficients left are not independent since they are
both related to the Gilbert damping coefficient $\eta$ and the
gyromagnetic factor $\Gamma$ .  The following relation holds
\cite{Revue}:

\begin{equation}
         L_{\theta 
        \theta} = -\alpha L_{\theta \varphi}
		 =  \frac{\rho_{0}^{F} \alpha \Gamma}{vM_{s}(1+ \alpha^{2})}
		 \label{Ltheta}
\end{equation}

where $\alpha = \eta \Gamma M_{s}$ is the normalized Gilbert damping coefficient.

In the absence of spin-injection $\Delta \mu^{e} = 0$ and there is no 
coupling between the currents. In that case, the well-known LLG equation of the 
ferromagnetic 
layer of volume $v$
is recovered by inserting the ferromagnetic chemical potential 
\cite{Mazur}

\begin{equation}
\mu^{F} = k_{B}T 
ln(\rho_{0}^{F}) + v V^{F}
\label{ChemPotFerr}
\end{equation}
into the expression of the ferromagnetic 
current density $\vec{j}_{0}^{F} = 
-\bar{\mathcal{L}} \vec{\nabla} \mu^{F}$ where $\bar{\mathcal{L}}$ is 
the 2x2 matrix of components $ \{ L_{\theta\theta}, L_{\theta \varphi},
L_{\varphi \theta}, L_{\varphi \varphi} \} = L_{\theta \theta }/\alpha \left 
(\{ \alpha, -1 , 1, \alpha \} \right ) $.
We have the expression:

\begin{equation}
	 \vec{j_{0}^{F}} v= 
			   - \frac{\rho_{0}^{F} \Gamma}{M_{s}(1+ \alpha^{2})}    \left \{ 
			   \vec{u}_{r}\times \left (
		 \vec{\nabla} V^{F}
		 + \frac{k_BT}{v}  \frac{\vec{\nabla} \rho_{0}^{F}}{ \rho_{0}^{F}} \right)
			   - \alpha \vec{u}_{r} \times \left[ 
			   \vec{u}_{r}\times \left (
			    \vec{\nabla} V^{F}  
			   +
			   \frac{k_BT}{v} \frac{\vec{\nabla} \rho_{0}^{F}}{ \rho_{0}^{F} }
			           \right) \right] \right \}
		   \label{flux}
\end{equation}

The LLG equation (that includes diffusion terms) is deduced immediately 
by dividing Eq. (\ref{flux}) with the density $\rho_{0}^{F}$, thanks to the relations $
 \rho_{0}^{F} \frac{d\vec u_{r}}{d t} = \vec{J}_{0}^{F}$ and $\vec{M} = 
 M_{s} \vec{u_{r}}$ \cite{Revue}. \\
    
  \begin{figure}
   \begin{center}
   \begin{tabular}{c}
   \includegraphics[height=7cm]{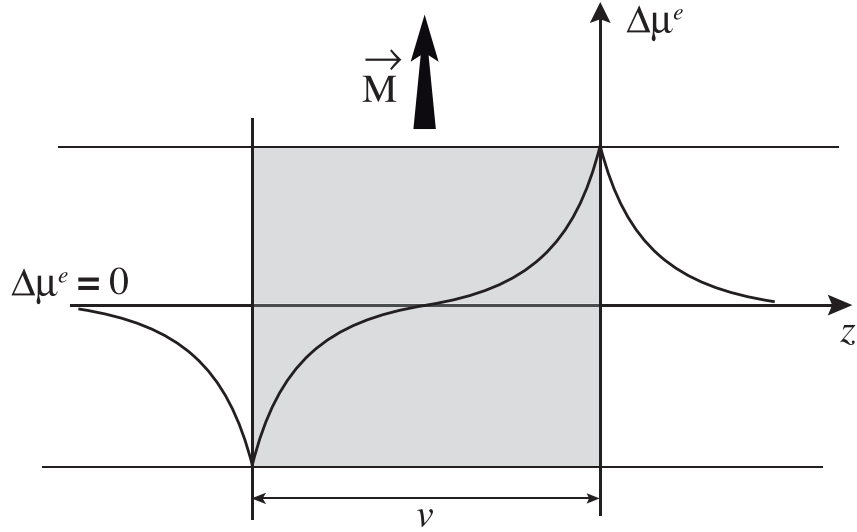}
   \end{tabular}
   \end{center}
   \caption[Interface] 
{ \label{fig:Int} Illustration of the spin-accumulation occurring in a
nanoscopic ferromagnetic layer with its two non-ferromagnetic
contacts: the profile of the electrochemical potential difference
$\Delta \mu^{e}$ is plotted as a function the spatial 
coordinate $z$.}
   \end{figure} 

      \begin{figure}
   \begin{center}
   \begin{tabular}{c}
   \includegraphics[height=7cm]{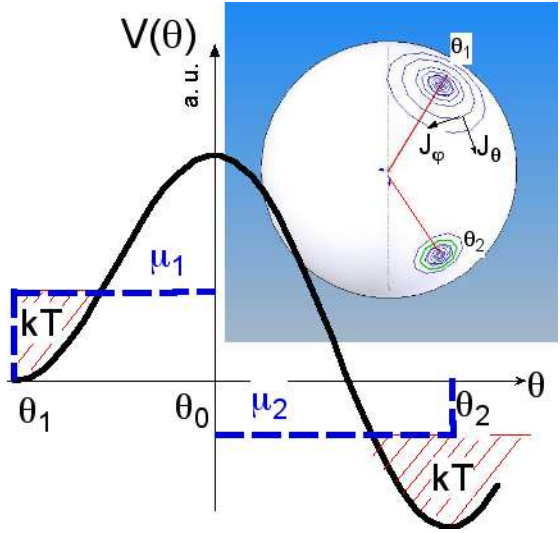}
   \end{tabular}
   \end{center}
   \caption[Potentiel] 
{ \label{fig:Pot} The configuration space of the ferromagnetic order 
parameter
is represented by a sphere of radius unity.  The double well shown is a 
projection over the plane
that contains the two equilibrium states at $\theta_{1}$ and $\theta_{2}$, and the 
top of the barrier at $\theta_{0}$.  The step approximation for the
chemical potential $ \mu(\theta)$ is plotted with the thermal fluctuations
sketched by the dashed area around the two minima.  }
   \end{figure}

   In the presence of spin injection, $\Delta \mu \ne 0$ and a 
   correction to the above LLG equation is expected, due to the
   phenomenological transport spin-transfer coefficient $l$, introduced in the
   Onsager matrix of Eq.  (\ref{Onsagermatrix}).  For the sake of
   simplicity, we only treat the coupling of {\it longitudinal} spin
   relaxation along the vector $\vec{u}_{\theta}$. 
   
Assuming $l $ constant, the longitudinal component of the total ferromagnetic 
current in the layer
$ \vec{\mathcal J}^{F}(\theta, \varphi, v) = \int_{v}  
\vec{j}^{F}(\theta,\varphi, z) dz $
has the form,

\begin{equation}
\mathcal J^{\theta F}(\theta, \varphi, v) = j_{0}^{\theta F}(\theta, 
\varphi) v
-  l \int_{v} \, \frac{\partial \Delta 
\mu^{e}}{\partial z}(\theta,z)dz
\label{Onsager1}
\end{equation}

The quantity 
$\beta \, \int_{v}
\frac{\partial \Delta \mu^{e}}{\partial z} dz$ is proportional to the
giant magnetoresistance $R_{GMR}$ generated at the interface 
\cite{ValetFert,Revue} (see Fig. 1 and \cite{condition}):

\begin{equation}
\beta \, \int_{v} \frac{\partial \Delta \mu^{e}}{\partial z} dz = 
8 e I R_{GMR}(\theta)
\label{GMR}
\end{equation}
where $J^{e}_{0} = I$ is the electric current injected in the junction
of section unity.  Since the giant magnetoresistance $R_{GMR}(\theta)$
depends on the angle between the incident spin-polarized current (e.g.
defined by the magnetization state of a second magnetic layer in a
spin-valve structure) and the ferromagnetic layer, it depends on the state of
the ferromagnetic layer, i.e. the position in the $\gamma$-space. 
Eq.  (\ref{Onsager1}) shows that the total ferromagnetic current is
simply related to the gradient of a total chemical potential
$\mu^{F}_{tot}$:

\begin{equation}
\vec{\mathcal J}^F =-\bar{\mathcal{L}} \vec{\nabla} \mu^{F}_{tot}
\label{CurrentTot}
\end{equation}

The {\it total chemical potential} 
$\mu^{F}_{tot}$ writes: 

\begin{equation}
\mu^{F}_{tot}(\theta) = \mu^{F}(\theta) +
\frac{l}{L_{\theta \theta}} \, \mathcal{V}^{e}(\theta)
\label{mutot}
\end{equation}

where the {\it electrospin chemical potential} $\mathcal{V}^{e}(\theta)$  is given by an integration
over the length of the ferromagnet in the direction $z$, and over the
angle $\theta $:

\begin{equation}
\mathcal{V}^{e}(\theta) = \frac{1}{e} \int_{\theta}
\int_{v} \frac{\partial \Delta \mu^{e}}{\partial z} (z,\theta') dz \,
d\theta' = \frac{8 e I}{\beta}  \int_{\theta}R_{GMR}(\theta') d\theta'
\label{M} 
\end{equation}

Inserting the {\it ferromagnetic chemical potential} Eq. 
(\ref{ChemPotFerr}) into Eq.  (\ref{mutot}) yields

\begin{equation}
\mu^{F}_{tot}(\theta) = kT \, ln(\rho_{0}^{F} \left (\theta) 
\right ) + \frac{l}{L_{\theta \theta}} \,
\mathcal{V}^{e}(\theta) + vV^{F}(\theta)
\label{muTot}
\end{equation}

In order to deal with the density of the total system $\rho_{tot}^{F}$, the
contribution of the spin-dependent scattering is included in the logarithm :

$\mu^{F}_{tot}= kT \, ln(\rho_{tot}^{F}) + vV^{F}$ where

\begin{equation}
\rho^{F}_{tot}(\theta) = \rho_{0}^{F}(\theta)
.e^{\frac{l}{kT L_{\theta \theta}} \, \mathcal{V}^{e}(\theta)}
\label{rhoTot}
\end{equation}
Consequently, the expression of the ferromagnetic-electrochemical potential
takes the same form as in the case of a ferromagnetic chemical
potential without spin-injection, but with a modified density (that
depends on the spin dependent scattering process).

The total ferromagnetic current, given by Eq. (\ref{CurrentTot}), 
writes now:

\begin{equation}
	 \vec{\mathcal J^{F}}= 
			   - \frac{\rho_{tot}^{F} \Gamma}{M_{s}(1+ \alpha^{2})}   \left \{ 
			   \vec{u}_{r}\times \left (
		 \vec{\nabla} V^{F}
		 + \frac{k_BT}{v}  \frac{\vec{\nabla} \rho_{tot}^{F}}{ \rho_{tot}^{F}} \right)
			   - \alpha \vec{u}_{r} \times \left[ 
			   \vec{u}_{r}\times \left (
			    \vec{\nabla} V^{F}  
			   +
			   \frac{k_BT}{v} \frac{\vec{\nabla} \rho_{tot}^{F}}{ \rho_{tot}^{F} }
			           \right) \right] \right \}
		   \label{fluxTot}
\end{equation}

Once again, the corresponding {\it generalized LLG equation} is
directly obtained from the expression $\frac{d\vec{u}_{r}}{dt} =
\frac{\vec{\mathcal J}_{tot}^{F}}{\rho^{F}_{tot}}$.  The equation can be rewritten
with introducing the "pseudo" effective field $\vec{\tilde H}_{eff} =
- \vec{\nabla} V^{F} + \frac{k_{B}T}{v} \frac{\vec{\nabla}
\rho_{tot}^{F}}{\rho_{tot}^{F}} $, so that the Generalized LLG
equation takes the usual form:

\begin{equation}
\frac{d\vec{u}_{r}}{dt} =   \frac{\Gamma}{M_{s}(1 + \alpha^{2})} \left \{ 
			   \vec{u}_{r}\times \vec{\tilde H}_{eff}
			   - \alpha \vec{u}_{r} \times \left[ 
			   \vec{u}_{r}\times \vec{\tilde H}_{eff} \right] \right \}
\label{GenLLGDEUX}
\end{equation}

This equation has the same form than that of the LLG equation without
spin-injection (note that it is mainly a consequence of the approximation of
longitudinal coupling only, disregarding precessional coupling), but
the diffusion part of the effective field is modified through the
correction of the density $\rho_{tot}^{F} = \rho_{0}^{F}
e^{\frac{l}{k_{B}T L_{\theta \theta}} \mathcal{V}^{e}} $ :

\begin{equation}
	k_{B}T \left ( \frac{\vec{\nabla} \rho_{tot}^{F}}{\rho_{tot}^{F}}
	- \frac{\vec{\nabla} \rho_{0}^{F}}{\rho_{0}^{F}} \right ) =
	\frac{l}{L_{\theta \theta}} \frac{ \partial
	\mathcal{V}^{e}}{\partial \theta} \vec{u}_{\theta}
	\label{correcLLG}
\end{equation}

the
correction due to spin-injection $\frac{8le}{\beta k_{B}T L_{\theta
\theta} v} R_{GMR}I $ is consequently a correction to the diffusion
term that is proportional to the GMR and to 
the injected current. 
However, this diffusion term accounts for fluctuations \cite{Brown}. 
The effect of the diffusion terms cannot be taken into account in the
quasi-static hysteresis loop (i.e. for vanishing temperature or 
infinite measurement times) as a
deterministic effective field (this point is discussed in
the references \cite{Moi,Revue}).  A deterministic correction to the
reversible part of the hysteresis (the quasi-static states) is hence
not expected here.  In contrast, the effect of this correction is
considerable in the activation regime or in ferromagnetic resonance near
equilibrium, described in the next sections.

\subsection{Spin-injection correction to the activation 
process}

What is the consequence of the diffusion correction Eq. 
(\ref{correcLLG}) in the activation regime of magnetization reversal? 
This question is investigated below for large time scales (beyond
nanosecond time scales, or "high barriers"), with the corresponding
activation process, i.e. the so-called N\'eel-Brown relaxation
\cite{SpinTorque}.  In the activation regime, the effect of the
precession can be neglected:  the gradient $\vec{\nabla}$ and the
divergence operator can be reduced to the scalar derivative
$\frac{\partial}{\partial \theta}$.

  
 The ferromagnetic potential $V^{F}$ has a double well structure 
 (Fig. 2),
 with the two minima $\theta_{1}$ and $\theta_{2}$, and a maximum at
 $\theta_{0}$.  The description of the activation process is based on
 the high barrier approximation under which the 
 ferromagnetic current becomes a
 step function in the $\theta$ space:
 
 \begin{equation}
\mathcal J^{F}(\theta,t) = \mathcal J^{F}(t) 
            \left [ 
            \Theta (\theta-\theta_1) - \Theta (\theta-\theta_2) 
			\right ]
\label{Jconstante}
\end{equation}
where $\Theta(\theta)$ is the Heaviside step function.

The chemical potential can also be approximated by a step function that
takes the value of the equilibrium states in the left ($\theta_{1}$)
or in the right ($\theta_{2}$) side of the potential barrier
in the ferromagnetic configuration space (Fig.  2) :
$\mu^{F}(\theta,t)=\mu^{F}(\theta_1,t) \Theta(\theta_0-\theta)
+\mu^{F}(\theta_2,t) \Theta(\theta-\theta_0)$.  Using now this 
expression, the density function $\rho_{tot}^{F}(\theta)$
of the ferromagnet in configuration space writes:

\begin{equation}
	\rho_{tot}^{F}(\theta, t) = \rho_{tot}^{F}(\theta_{1},t)
	e^{-\frac{v \left ( V^{F}(\theta) - V^{F}(\theta_{1}) \right
	)}{k_{B}T}} \Theta (\theta_{0}-\theta) +
	\rho_{tot}^{F}(\theta_{2},t) e^{-\frac{v \left ( V^{F}(\theta) -
	V^{F}(\theta_{2})\right )}{k_{B}T}} \Theta (\theta-\theta_0)
	\label{density}
	\end{equation}

The ferromagnetic current is related to the gradient of the 
generalized chemical potential through 
our fundamental relation Eq. (\ref{CurrentTot}) :
$\mathcal J^{F}(\theta,t) =- L_{\theta \theta}
             \frac {\partial{\mu_{tot}^{F}(\theta,t)}}
                   {\partial{\theta}}$. This equation
can be written into the more convenient form
\begin{equation}
\mathcal J^{F}(\theta,t) = -D_{tot}^{F}(\theta) 
\, e^{-\frac{vV^{F}(\theta)}{k_B T}} \, 
\frac{\partial e^{\frac{\mu^{F}_{tot}(\theta,t)}{k_BT}}}{\partial \theta }
\label{Onsager2}
\end{equation}

with the diffusion coefficient 

\begin{equation}
D_{tot}^{F} (\theta) \equiv \frac{kT L_{\theta 
\theta}}{\rho_{tot}^{F}(\theta)} = D_{0}^{F} .e^{-\frac{l
\mathcal{V}^{e}(\theta)}{k_{B} T L_{\theta \theta}}}
\label{diffcoefTot}
\end{equation}

where the second equality is deduced from Eq.  (\ref{rhoTot} ) and the
parameter $ D_{0}^{F}= L_{\theta \theta}.k_B T/\rho_{0}^{F}$
(dimension of angle per unit of time) is the usual ferromagnetic
diffusion coefficient that is constant.

The activation process is described by a rate equation (see Eq. 
(\ref{JfinalDEUX}) below), e. i. a contracted description that is obtained by
performing a reduction of the continuous internal variable $\theta$
over the equilibrium states $\theta_{i}$ ($i=\{1,2 \}$). 

The total flow has a zero divergence current density $div \mathcal
J^{F} = 0$.  The system is quasi stationary and the total current is
$\mathcal I = 2 \pi sin(\theta) \mathcal J(t) $.  Eqs. 
(\ref{Jconstante}) and Eqs.  (\ref{Onsager2}) can be integrated over
the measure $exp(vV/k_{B}T) d \theta$ to give:

\begin{equation}
\mathcal I \int \frac{e^{\frac{vV^{F}(\theta)}{k_{B}T}}}{2 \pi sin \theta} 
            \left [ 
            \Theta (\theta_{1}-\theta) - \Theta 
            (\theta_2 - \theta) 
			\right ] d\theta = 
			- \int D_{tot}^{F}(\theta) 
\, e^{-\frac{vV^{F}(\theta)}{k_B T}} \, 
\frac{\partial e^{\frac{\mu^{F}(\theta)}{k_BT}}}{\partial 
\theta } d\theta
\label{Integr}
\end{equation}

so that the total current writes \cite{RqueDistrib}:

\begin{equation}
\mathcal I = D_{tot}^F(\theta_{0})
\frac{e^{\mu(\theta_{2})/k_{B}T}-e^{\mu(\theta_{1})/k_{B}T}}{\int^
{\theta_2}_{\theta_1} \frac{e^{vV^{F}(\theta)/k_BT}}{2 \pi \, sin(\theta)} \,d\theta}
\label{JActiv}
\end{equation}
	
Defining the number of representative points near equilibrium by
$n(\theta_{i}) = \int_{-\epsilon}^{\theta_{i} + \epsilon} \rho(\theta') 2 \pi
sin(\theta') d \theta'$, the density in the double well potential Eq. 
(\ref{density}) leads to the expressions:

\begin{equation}
e^{\mu(\theta_{1})/k_{B}T} = 
\frac{n_{tot}(\theta_{1})}
              {2 \pi \int_{-\epsilon}^{\theta_{0}} e^{-vV^{F}(\theta)/k_{B}T}
             sin(\theta) d\theta }
			  \nonumber
\end{equation}
\begin{equation}
e^{\mu(\theta_{2})/k_{B}T} =
\frac{n_{tot}(\theta_{2})}{
2 \pi \int_{\theta_{0}}^{\theta_{2} + \epsilon}
e^{-vV^{F}(\theta)/k_{B}T} sin(\theta) d \theta }
\label{Muni1}
\end{equation}

where $\epsilon$ is a real number. Inserting the above equation into Eq. 
(\ref{JActiv}) leads to the generalized rate equation:

\begin{equation}
\mathcal{I} = \dot n_{tot}(\theta_{1}) = -\dot n_{tot}(\theta_{2}) = 
\frac{n_{tot}(\theta_{1})}{\tilde \tau_{1 \rightarrow 2}} -
\frac{n_{tot}(\theta_{2})}{\tilde \tau_{2 \rightarrow 1}}
\label{JfinalDEUX}
\end{equation}
Using the steepest descents approximation \cite{RqueSteepest} for the
three integrals present in Eq.  (\ref{JActiv}) after inserting
Eqs.  (\ref{Muni1}), the total relaxation times write:

\begin{equation}
\tilde \tau_{i \rightarrow i\pm 1}^{-1} = D_{tot}(\theta_{0}) 
\frac{sin(\theta_{0})}{sin(\theta_{i})}
\frac{
v\sqrt{|(V^{F})''(\theta_{0})|| (V^{F})''(\theta_{i})|}}{2 \pi kT}
e^{\frac{vV^{F}(\theta_{i}) - vV^{F}(\theta_{0})}{k_{B}T}}
\label{Arrhenius}
\end{equation}

More explicitly, this generalized rate equation as the form of the
usual N\'eel -Brown relaxation rates $\tau_{i \rightarrow i\pm 1}$,
with an exponential correction expressed in terms of the {\it electrospin}
chemical potential $\mathcal{V}^e(\theta_{i})$:

\begin{equation}
\dot n_{tot}(\theta_{1}) = -\dot n_{tot}(\theta_{2}) = \frac{n(\theta_{1})}{
\tau_{1 \rightarrow 2}} 
e^{- \frac{l \left ( \mathcal{V}^{e}(\theta_{1}) - 
\mathcal{V}^{e}(\theta_{0}) \right ) }{k_{B}T \, L_{\theta \theta} }} -
\frac{n(\theta_{2})}{\tau_{2 \rightarrow 1}}
e^{- \frac{l \left ( \mathcal{V}^{e}(\theta_{2}) - 
\mathcal{V}^{e}(\theta_{0}) \right)}{ k_{B}T \, L_{\theta \theta}}}
\label{JfinalTROIS}
\end{equation}
where $\mathcal{V}^{e}(\theta) = 8eI \int_{0}^{\theta} 
R_{GMR}(\theta') d\theta' / \beta$.
Accordingly, the N\'eel-Brown activation law is still valid
under current injection, with a correction
that can be added to the potential energy barrier. 
The total potential energy including the contribution of the 
spin-injection writes :

\begin{equation}
V_{tot}(\theta) = V^{F}(\theta) + \frac{8l eI}{\beta
L_{\theta \theta} } \int_{\theta} R_{GMR} (\theta') d\theta'
\label{Barrier_{eff}}
\end{equation}
The correction to the 
energy barrier is proportional
to the current I and to the GMR integrated over the 
magnetization states corresponding to the energy barrier height.  Eq. 
(\ref{JfinalTROIS}) shows that the process follows the N\'eel-Brown 
activation law 
with the relaxation times: 
$\tilde \tau = \tau_{0}\, e^{- \frac{\Delta V_{tot}}{k_{B}T}} $
where $\Delta V_{tot} = V_{tot}(\theta_{i}) - V_{tot}(\theta_{0})$,
$\theta_{i} = \{\theta_{1}, \theta_{2} \}$ and $\tau_{0}$ is the
usual waiting time (i.e. the prefactor in Eq.(\ref{Arrhenius}) ). 

In order to compare this analysis with experimental results performed 
on nanopillars, let us
assume a spin-valve structure with two ferromagnetic layers composed
of identical materials with $\beta \ge 0$, in which only two states along the anisotropy
axes are allowed.  One layer is fixed (the pinned layer) and the
states of the other (the "free" layer) are investigated.  The
magnetization states of the free layer are $\theta_{1} =0$
for the parallel configuration (P) and $\theta_{2} = \pi$ for
the ant parallel configuration (AP).  The GMR for the P state $R_{GMR}(0)
= 0$ corresponds to the reference configuration (no spin-flip).  The
AP configuration $R_{GMR}(\pi) = \Delta R$ corresponds to the
maximum GMR. If we take the most simple form for the angular 
dependence of the GMR
\cite{Dieny} $R_{GMR}(\theta) = \Delta R \, \left (1 - 
cos(\theta) \right )$, we have 
$\mathcal V^{e}(0) - \mathcal V^{e}(\pi/2)  = - \frac{4 I 
 \Delta R (\pi-2)}{ 
\beta }$  and 
$\mathcal V^{e}(\pi) - \mathcal V^{e}(\pi/2)  = + \frac{4 I 
\Delta R (\pi + 2)}{ 
\beta}$.

 Note that the GMR 
parameter $\Delta R$ is a function of $\beta$: expressed in terms of 
the spin diffusion length $l_{sf}$ it writes \cite{Revue} $\Delta R/R = 
\frac{\beta^{2}}{1-\beta} \frac{l_{sf}}{v}$ (where $v$ is the length 
of the layer). 

In conclusion, the injection of the current leads to suppress one
transition and to accelerate the other: the current provokes the
magnetization reversal from one configuration to the other, and the
transition depends on the current direction.  In the exemple ebove, the
current provokes the magnetization reversal from P to AP configuration
for positive current, and provokes the magnetization reversal from AP
to P configuration for negative current.  This is a sufficient
condition in order to accounts for the hysteresis loop of the
magnetization driven by the current.  In the general
case, both transitions are defined with the relaxation rate 
\begin{equation}
\tau_{i}
= \tau_{0} \, e^{- \frac{ \Delta V^{F} \pm \frac{c_{i} \, l \Delta R}{
 \beta L_{\theta \theta}} I}{2 k_{B} T} }
\label{ActivTot}
\end{equation}
with a asymmetry factor $c_{i}
\equiv \int_{\theta_{i}}^{\theta_{0}} R_{GMR}(\theta) d\theta / \Delta 
R $, and the coefficient $L_{\theta,\theta}$ is defined in Eq. 
(\ref{Ltheta})

( $c_{P} = (\pi - 2)/2$ and $c_{AP}= (\pi + 2)/2$ in the simple 
exemple given above).  The quasi-symmetry under both the permutation of the
magnetic configurations and the change of the current direction
observed experimentally is hence contained in the result expressed in
Eq.  (\ref{JfinalTROIS}).

From an empirical point of view, the expression often used in order to
fit the data introduces the critical current $I_{c}$ (measured at zero
external field and extrapolated at zero Kelvin), such that: $ \tilde
\tau = \tau_{0}\, e^{ - \frac{E_{a}}{kT} (1- \frac{I}{I_{c}})} $ where
$E_{a}$ is the anisotropy energy of the ferromagnetic layer under
consideration.  Result Eq.  (\ref{JfinalTROIS}) shows that the 
critical current is given by the expression:
\begin{equation}
I_{c} = -\frac{\Gamma \alpha E_{a}}{v M_{s} (1+ \alpha^{2}) c_{i}} 
\left ( \frac{\rho^{F}_{0}}{el} \right) \frac{\beta}{8 \Delta R}
\label{Icrit}
\end{equation}

The N\'eel-Brown law under current injection is measured in references
\cite{MSU2,Moi,Fabian,Buhrman}, with the typical asymmetry between the
two transitions of a factor 2 to 4.  The relation $I_{c} \propto
1/\Delta R$ is verified in reference \cite{MSU2}.  The proportionality
with $\beta$ is observed through the change of the sign while changing
the scattering anisotropy \cite{MSU3}.  Note that the phenomenological
results presented here can be generalized to tunnel junctions (see
e.g. the work of Schmidt and et al. in terms of spin injection in
magnetic semiconductors \cite{Molenkamp}).  In the case of tunnel
barrier a factor ten is typically gained
in the magnetoresistance $\Delta R$, so that the critical currents in  Eq. 
(\ref{Icrit}) are also decreased by a factor ten \cite{TJ}.

What is the value of the spin-transfer coefficient $l$?  In order to
compare with the ferromagnetic transport coefficient $L_{\theta
\theta}/\rho_{0}^{F} = \Gamma \alpha / (M_{s}(1 + \alpha^{2}))$ -
expressed as the inverse of an action $J^{-1}.  s^{-1}$ - the
phenomenological coefficient is compared in the same units:
$\frac{l}{e \rho_{0}^{F}} = -\frac{E_{a} \beta}{e 8 \Delta R I_{c}
c_{i}} \left ( \frac{L_{\theta \theta}}{\rho_{0}^{F}} \right )$.

Experiments performed on typical pseudo spin-valve systems show that
it is possible to switch the magnetization at zero external field in
both directions (AP to P or P to AP in the previous exempla) for
currents of the order of $\pm$ 1 mA. For such currents, the energy
transferred is of the order of 10 meV \cite{MSU2}: the measured
quantity is the slope of the points $1/I_{c}$ plotted
as a function of $\Delta R$ for the two transitions.  The anisotropy
energy $E_{a}$ is of the order of 0.1 $eV$.  The spin-transfer
coefficient $l/e$ is consequently of the order of $10^{-1} L_{\theta
\theta}$ to $10^{-2} L_{\theta \theta}$.

On the other hand, the activation experiments under current injection
allow to access directly to the spin-transfer coefficient through the
N\'eel-Brown law, without the need to measure the activation energy
$E_{a}$.  The quantity measured is the slope $s =
\partial(ln(\tau/\tau_{0}))/\partial I$.  Eq.  (\ref{Barrier_{eff}})
and Eq.  (\ref{ActivTot}) show that: $ \frac{l}{eL_{\theta \theta}} =
kT s \beta \left ( 8 e \int_{\theta_{eq}}^{\theta_{0}}
R_{GMR}(\theta')d \theta' \right )^{-1} \approx 0.1$.  The order of magnitude 
$10^{-1}$ is confirmed experimentally in references \cite{MSU2,Moi}, together with the factor 2
to 4 in the asymmetry.

 \subsection{Spin-injection correction to the fluctuations}
   
 Due to its diffusive nature, the correction produced by the
 spin-injection can hardly be observed on the reversible state of
 the hysteresis.  The above section shows that the activated
 irreversible jump of the magnetization is in contrast strongly
 modified by the spin-injection.  Beyond the activation process, the presence
 of supplementary ferromagnetic diffusion processes strongly affects
 another experimentally accessible parameter: the linear response of
 the ferromagnetic moment to magnetic field, spin-injection, and
 thermal excitations.  The response is then proportional to the
 fluctuations.
 
 The fluctuations 
 occurring near the quasi-static states in the double-well potential 
 can be analyzed from the 
 general fluctuation-dissipation theorem (FDT) in the $\gamma $-space 
 \cite{RubiFDT}.
 The density $\tilde \rho^{F}$ is subjected to random fluctuations that are 
 introduced through a {\it random current}  $\mathcal J_{r}^{F}$, 
 which satisfies FDT: 
 
 \begin{equation}
 \langle \mathcal J^{F}_{r}(\theta , t) \mathcal J_{r}(\theta',t') \rangle = 
 2 D_{tot}^{F}(\theta)  \langle 
 \rho_{tot}^{F} (\theta,t)\rangle \delta(\theta - \theta') \delta(t-t') 
 \label{FDT1}
 \end{equation}
 
 The variation of density is
 now corrected by the presence of the fluctuation current:
 
 \begin{equation}
\frac{\partial}{\partial t} \rho_{tot}(\theta,t) = - \mathcal J^{F}(\theta,t) - 
\mathcal J_{r}^{F}(\theta,t)
\label{n1n2BIS}
\end{equation}
Applying step by step the method described above for the rate
equation (Eqs.  (\ref{Integr}) to (\ref{JfinalDEUX})) to the Eq. 
(\ref{n1n2BIS}) , the following expression of the fluctuations is
obtained (see reference \cite{RubiFDT}):

\begin{equation}
\langle \mathcal I_{r}(t) \mathcal I_{r}(t') \rangle = 
\frac{ \langle n_{tot}^{F} \rangle (\theta_{1})}{\tilde \tau_{1 \rightarrow 
2}} - 
\frac{\langle n_{tot}^{F}\rangle (\theta_{2}) }{\tilde \tau_{2 
\rightarrow 1}} 
\label{Jfluct}
\end{equation}

where the relaxation times $\tilde \tau$ are that previously defined. 
This expression has not the usual form of a FDT which means that this
theorem, strictly valid when fluctuations take place around
equilibrium states, is not fulfilled.  The theorem is restored near an
equilibrium state $ \langle n_{tot}^{F} \rangle ^{eq} = \langle
n_{tot}^{F} \rangle (\theta_{1})$ or $ \langle n_{tot}^{F} \rangle
^{eq} = \langle n_{tot}^{F} \rangle (\theta_{2})$ because transitions
from one equilibrium state to the other are neglected.  We obtain from
Eq.  (\ref{Jfluct}) $\langle \mathcal I_{r}(t) \mathcal I_{r}(t')
\rangle^{eq} = \frac{\langle n_{tot}^{F} \rangle ^{eq}}{\tilde
\tau_{eq}} \delta(t-t')$.  This last expression is valid in the cases
of linear ferromagnetic resonance experiments, i.e. in a situation
where {\it the current is well below the critical current} $I_{c}$
defined in Eq.  (\ref{Icrit}) \cite{DienyRes}.  A more complicated
behavior (highly non-linear) should be expected for strong
excitations near or beyond the critical current $I_{c}$ in order to 
interpret the non-linear resonance experiments \cite{STReson}.

In conclusion, in the case of linear ferromagnetic resonance (FMR) 
measured below $I_{c}$ and
observed close to one equilibrium state ($\theta_{1}$ or $\theta_{2}$), a
correction to the response of the ferromagnet is expected, that takes
the same form as that calculated for the activation process:

\begin{equation}
\langle \mathcal I_{r}(t) \mathcal I_{r}(t') \rangle^{eq} = 
2 D_{0}^{F} \, \frac{\langle n_{0}^{F} \rangle}{\tau_{eq}} 
exp \left ( - \frac{l(\mathcal{V}^{e}(\theta_{i}) - 
\mathcal{V}^{e}(\theta_{0}))}{k_{B}T \, L_{\theta \theta}} \right )  
\delta(t-t')
\end{equation}

 The behavior expected is then surprisingly similar to that predicted
 in the case of the activation process, except that it holds for the
 amplitude of the linear response and not for the transition rates.  For 
 $\beta \ge 0$, we expect an
 exponential increase (resp.  suppression) of the response in the
 AP state with a positive (resp.  negative) current, and an
 exponential increase (resp.  suppression) of the response in the
 P state with a negative (resp.  positive) current.  This
 highly specific characteristic is in agreement with that observed
 experimentally in the context of FMR measurements under spin-injection
 below critical current $I_{c}$, i.e. a situation in which the
 magnetization is close enough to equilibrium states (see results
 presented in reference \cite{DienyRes}).
 
 \subsection{Link with microscopic theories}
Before concluding, a last question must be invoked about the relation
between the model presented here and the microscopic theories of
spin-transfer torque \cite{Slon,Berger,Miltat,ListTheo}.  The
phenomenological transport coefficient $l$ (and of course the known
transport coefficient $L_{\theta \theta}$ and $\beta \sigma_{0}$)
could formally be defined from the relevant Hamiltonian expression
with the help of projection-operator formalisms \cite{Fick}, or any
other techniques \cite{Kambersky,Montigny} that lead to the coupled
stochastic transport equations of the spin-polarized current and the
ferromagnetic order parameter in the corresponding configuration
space.  The difficulty is to manipulate on an equal footing a
microscopic degree of freedom, the spin of conduction electrons, and a
collective variable, the magnetic order parameter.  This task is far
beyond the present report, but it is possible to gain some insight
with dimension considerations.  The physical mechanism proposed here
for spin-transfer is based on the spin-injection only, that is
responsible for the supplementary diffusion effect of the
magnetization through the modification of the local densities of
magnetic moments (redistribution of spins).  This redistribution of
spins between the electric sub-system and the magnetic sub-system is
governed by specific spin-flip scattering mechanisms (or
spin dependent creation-anhilation mechanisms).  The microscopic
approach would define the relevant mechanism and deduce the typical
spin-transfer relaxation time $\tau_{tr}$.  The relation between
mesoscopic and microscopic approaches can consequently be invoked
through the relation between the correction of the diffusion constant
$\delta D_{tot}^{F}$ (expressed in dimension of angle per unit of
time) and the relaxation time: $\delta D_{tot}^{F} \propto \delta
n/\tau_{tr}$, where $\delta n$ is the amount of spins transferred from
the electric sub-system to the ferromagnet.
 
 In the approach proposed by Berger \cite{Berger} in a pioneering
 work, a spin-transfer process at the interface is described at the
 electronic level by a typical spin relaxation time $\tau_{sd}$
 calculated from the s-d exchange Hamiltonian.  If we assuming that
 the relevant spin-flip relaxation is governed by this mechanism 
 $\tau_{tr} = \tau_{sd}$, we
 would have $\delta D_{tot}^{F} \propto 1/\tau_{sd}$ under the
 relevant hypotheses.  In that case, the spin-transfer described here
 in terms of diffusion process would be a consequence, in parallel to
 GMR effects, of the s-d exchange interaction occuring at the
 microscopic level.
 
\subsection{Conclusion}
A description of spin-transfer has been proposed at the mesoscopic
level, based uniquely on the spin-injection mechanism occuring at the
junction with a ferromagnet.  The spin-accumulation at the interface
leads to a local change of the density of magnetic moments in the
corresponding configuration space.  The gradient of density generates
a diffusion process of the ferromagnetic order parameter, which is
responsible for the magnetization reversal.

The spin-injection has been described at the interface of a
ferromagnet by means of the usual two-conduction channel model which
simplifies the spin-polarized current and the giant magnetoresistance
analyses.  The dynamics of the ferromagnet is treated by means of an
out-of-equilibrium mesoscopic model with a ferromagnetic current defined
in the configuration space of uniform magnetic moments.  The coupling
between the two currents is introduced through a new phenomenological
Onsager cross-coefficient $l$ that
accounts for the fact that the spins carried by the electric charges
in the normal space contribute also to the transport of ferromagnetic
moments in the ferromagnetic configuration space.  It as been shown
that the correction to the LLG equation that governs the dynamic of
the magnetization comes from a diffusion term.  We have found that the
N\'eel-Brown activation law is still valid, with a correction to the
barrier height that is proportional to the integral of the giant
magnetoresistances over the ferromagnetic states, from the equilibrium
to the top of the barrier.  The expression of the critical current
$I_{c}$ is given as a function of GMR, the damping factor and the new
spin-transfer cross-coefficient $l$.  Furthermore, the correction to the fluctuations
is shown to be analogous to that of the activation and is also
expressed as an exponential term.  These results are consistent with
the results obtained experimentally in quasi-static modes (hysteresis
loops as a function of the current), in the activation regime under
spin-injection, and in linear resonance experiments under spin-injection with
$I \le I_{c}$.


\begin{references}
\bibitem{RqueDistrib} According to 
the expression of the chemical potential $\mu(\theta)$ in terms of Heaviside 
distributions, the 
derivative in the sense of the distribution of $exp(\mu/kT)$ is
	$\frac{\partial e^{\mu(\theta)/k_BT}}{\partial 
	\theta } = \left ( e^{\mu(\theta_{1})/k_{B}T} - 
	e^{\mu(\theta_{2})/k_{B}T} \right) \delta(\theta_{0} - \theta)$
where $\delta$ is the Dirac distribution.
\end{references}
\end{document}